\documentclass[notitlepage,onecolumn,showpacs,11pt]{revtex4-1}
\usepackage[utf8]{inputenc}
 \usepackage{tikz}
\usepackage{graphicx}
\usepackage{amsmath}    
\usepackage{amsfonts}
\usepackage{fullpage}
\usepackage{subcaption}
\usepackage{soul}
\usepackage{xcolor}

\DeclareMathOperator{\Tr}{Tr}
\DeclareMathOperator{\sech}{sech}

\begin{document}

\title{Exploring the extent of validity of quantum work fluctuation theorems in the presence of weak measurements}

\author{Sourabh Lahiri}
\affiliation{Department of Physics, Birla Institute of Technology Mesra, Ranchi, Jharkhand 835215, India}

\author{Subhashish Banerjee}
 \affiliation{Indian Institute of Technology Jodhpur, Jodhpur 342037, India}

\author{A. M. Jayannavar}
 \affiliation{Institute of Physics, Bhubaneswar, Odisha 751005}

\begin{abstract}
 Quantum work fluctuation theorems are known to hold when the work is defined as the difference between the outcomes of projective measurements carried out on the Hamiltonian of the system at the initial and the final time instants of the experimental realization of the process. A recent study showed that the theorem breaks down if the measurement is of a more general nature, i.e. if a \emph{positive operator valued measurement} is used, and the deviation vanishes only in the limit where the operators become projective in nature. We study a simple two-state system subjected to a unitary evolution under a Hamiltonian that is linearly dependent on time, and verify the validity of the above statement. We further define a weak value of work  and show that the deviation from the exact work fluctuation theorems are much less in this formalism. 
\end{abstract}

\maketitle


\newcommand{\av}[1]{\langle #1 \rangle}
\newcommand{\alert}[1]{{\color{red} #1}}

\section{Introduction}

The field of nonequilibrium statistical mechanics has received a huge impetus after the discovery of the so-called \emph{fluctuation theorems} 
more than two decades ago \cite{harbola,jarzynski,hanggi,seifert,anders}. These theorems are distinctly different from their predecessors such as the linear response theory \cite{rmplinearresponse}, in that they remain valid even in a highly nonequilibrium state of the system. In fact, the linear as well as nonlinear response coefficients can be obtained by making use of these theorems \cite{gaspard2007}. The relations have been proved under a variety of conditions both for classical and quantum systems \cite{hanggi,seifert,jarzynski2010,jarzynski2007,jarzynski2006,seifert2005,seifert2005a,hanggi2009,hanggi2010}. One of these relations, known as the Crooks' Theorem \cite{crooks1,crooks2} pertains to the work $W$ done on the system, along the forward and along the reverse processes (each beginning in a state of thermal equilibrium):

\begin{align}
 \frac{P_f(W)}{P_r(-W)} &= e^{\beta(W-\Delta F)},
\end{align}
where $P_f(W)$ and $P_r(-W)$ are the probability distributions of work done in the forward and reverse processes, while $\Delta F$ is the change in the equilibrium free energy in the forward process. $\beta$ is the inverse temperature of the thermal environment. A famous corollary of the above theorem is the Jarzynski equality \cite{jarzynski1}, which is given by

\begin{align}
 \left< e^{-\beta W}\right> &= e^{-\beta\Delta F}
\end{align}
 The Jarzynski relation in a closed driven	quantum system has been derived  by introducing an initial and final projective 	measurment of the system energy  \cite{monnai}. Work in this context is a two-point quantity obtained by calculating the difference 	between the initial and final energy of the system \cite{talknerwork}. See also, in this context \cite{paz}, where work on a quantum system was shown to be measured by performing a generalized quantum measurement at a single time. In the context of open quantum systems, the ambient environment, also called the reservoir, needs to be taken into account \cite{weiss,sb}. In that case, the Jarzynski and Crooks relation have been studied from a variety of perspectives \cite{mukamel,mallick,quan}. Experimental schemes have been proposed to verify the quantum non-equilibrium fluctuation relations \cite{vedral}. These led to development of thermodynamics  of systems at the mesoscopic level and to the foundation of the field of stochastic thermodynamics \cite{sekimoto1998,siefertstochastic,sekimoto}. In turn, this has made a positive impact on the rapidly developing field of quantum thermodynamics \cite{kosloff,lutz,jayan,george}.
Both the Jarzynski and Crooks relations provide a way to measure free energy from nonequilibrium work measurements, so that one need not make the process reversible in order to do so.

The classical FTs rely on the stochastic definitions of thermodynamic quantities like work, heat or internal energy as prescribed in \cite{sekimoto1998,sekimoto}. The quantum definitions are more involved, since work was shown not to be a quantum observable \cite{talknerwork}. Nevertheless, a two-point measurement scheme seemed to provide the appropriate definition of work done on a quantum particle. It was argued in \cite{yi-kim} that if a projective measurement of Hamiltonian is performed on the combined system consisting of the system of interest and its environment, both at the beginning and at the end of the experiment, then the difference between the eigenvalues would physically correspond to the work done on the system of interest. 

Now, an experimental measurement of any quantum observable is in general \emph{not} a projective measurement. Rather, it is some form of a \emph{positive operator-valued} measurement (POVM) \cite{brun2002}.  A POVM of particular interest is called a weak measurement, where the disturbance to the system state is minimal. This is in stark contrast to the commonly used concept of projective measurement, where the state of the system collapses to one of its eigenstates after the measurement. 
 The weak measurements can be formulated using the pre- and post-selected quantum systems \cite{aharanov} as well as in terms of measurement operator formalism \cite{brun}. The measurement operator approach provides a new way to handle weak as well as strong measurements (generated by projective measurements).

In this article, we focus on the validity of the Crooks theorem and the Jarzynski equality when the measurement is weak in nature.  An attempt in this direction was made in \cite{venkatesh}, where the effect of generalized measurements on fluctuation relations was studied. Here we complement this study by the use of weak measurement operators, using the measurement operator approach \cite{brun} that do not require post-selection.  The points of contact as well as differences with this study are highlighted. To the knowledge of the authors, such checks have not been performed till date on the  fluctuation theorems. Further, the formalism facilitates the deduction of the usual case due to projective measurements easily. If a connection between the equilibrium free energy change and the outcomes of the weak measurements is available, it would therefore be of great practical relevance. 

We first discuss briefly about the known relations for projective measurements, and then proceed to extend the treatment to weak measurements.  For analytical convenience, we choose the protocol such that the Hamiltonian of the system has a linear dependence on time, although the results can be explained using very general physical reasonings (see later) that should remain valid even for more complex protocols.

In section \ref{sec:review_projective}, we discuss the notations and the model, and provide a brief derivation of the Jarzynski Equality for isolated quantum systems subjected to projective measurements, which is a known result. Section \ref{sec:weak_operators} provides an introduction to the weak measurement operators. In section \ref{sec:comparison}, we put our analytical results in perspective with the conclusions of \cite{venkatesh}. Section \ref{sec:new_results} discusses further generalizations and the underlying assumptions as contrasted with those of \cite{venkatesh}. 
We then make our Conclusions.

\section{Fluctuation Theorems for isolated quantum systems in presence of projective measurements}
\label{sec:review_projective}

\subsection{Description of the model}

\begin{figure}[!h]
 \centering
 \includegraphics[width=0.3\linewidth]{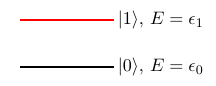}
 \caption{Diagram showing the two non-degenerate states of the system.}
 \label{fig:two_states}
\end{figure}

Let us consider a two-state system with the ground state of energy $\epsilon_0$ and the excited state of energy $\epsilon_1$, as shown in figure \ref{fig:two_states}.


\noindent  The system is initially (at $t=0$) in thermal equilibrium with a heat bath of temperature $T$. The two states are represented by
\begin{align}
 |0\rangle &= \begin{pmatrix}
               1 \\ 0
              \end{pmatrix};
              \hspace{1cm}
|1\rangle = \begin{pmatrix}
               0 \\ 1
              \end{pmatrix}.              
\end{align}

\noindent The initial Hamiltonian is 
\begin{align}
 H= \epsilon_0|0\rangle \langle 0| + \epsilon_1|1\rangle \langle 1| = \begin{pmatrix}
                                                                       \epsilon_0 & 0\\
                                                                       0 & \epsilon_1
                                                                      \end{pmatrix}.
\end{align}

\noindent The initial density matrix is given by 
\begin{align} \label{initial state}
 \rho(0) = \frac{1}{Z(0)}e^{-\beta H} = \frac{1}{Z(0)}\begin{pmatrix}
                        e^{-\beta \epsilon_0} & 0\\
                        0 & e^{-\beta \epsilon_1}
                       \end{pmatrix},
\end{align}

where \[
Z(0) = e^{-\beta \epsilon_0}+e^{-\beta\epsilon_1}.
\]

Once our protocol begins (we switch on the external drive $\lambda(t)$, so that the Hamiltonian becomes explicitly time-dependent), we detach our system from the heat bath, so that its subsequent evolution is unitary.

We now measure the energy eigenvalue $\epsilon_i$ at $t=0$, allow its unitary evolution till $t=\tau$, and make a final measurement of its energy eigenvalue $\epsilon_j$ (given: $i,j=0,1$). The change in energy $\epsilon_j(\tau)-\epsilon_i(0)$ must be the work done on the particle, since there is no dissipation of heat. 

Now, the probability distribution of work in this \emph{forward} process is given by
\begin{align}
 P_f(W) &= \sum_{i,j} \delta[W-(\epsilon_j-\epsilon_i)] P(j|i)p_{i}(0).
 \label{Pf}
\end{align}
Here, $p_i(0)=e^{-\beta\epsilon_i}/Z(0)$ is the initial thermal distribution of the state $|i\rangle$, $i=0,1$. The transition probability from state $|i\rangle$ to state $|j\rangle$ is given by $P(j|i)$. 

The \emph{reverse} process is defined  as one in which the system is again brought in thermal equilibrium with the bath at time $t=\tau$ (when the value of the external parameter is $\lambda(\tau)$), so that the initial density matrix $\tilde\rho(0)=e^{-\beta H(\tau)}/Z_\tau$ is similar to that in the forward process, but involving energy eigenvalues corresponding to the final form of the Hamiltonian $H(\tau)$.
\subsection{Jarzynski Equality for projective measurements}
The derivation for Jarzynski Equality (JE) when projective measurements are performed at the beginning and the end of the process is well-known \cite{talknerwork}. We briefly outline the method below.

Let us define the projection operators $\Pi_0\equiv |0\rangle\langle 0|$ and $\Pi_1 = |1\rangle\langle 1|$. The unitary evolution of the state for a time $t$ is given by the operator $U(t)$.

  In unitary evolution, $P(j|i)=P(i|j)$. Also, since the probability distributions $p_i(0)$ and $p_j(\tau)$ are Boltzmann distributions, we must have
\[
 \frac{p_i(0)}{p_j(\tau)} = e^{\beta(\epsilon_j-\epsilon_i)-\Delta F},
\]
so we can write \eqref{Pf} as
\begin{align}
 P_f(W) &= \sum_{i,j} \delta[W-(\epsilon_j-\epsilon_i)] P(j|i)p_{i}(0)\nonumber\\
 &= \sum_{i,j} \delta[W-(\epsilon_j-\epsilon_i)] \Tr[\Pi_j U(\tau)\Pi_i \rho(0)\Pi_i U^\dagger(\tau) \Pi_j].
 \label{PfW}
\end{align}
The characteristic function, given by the Fourier transform of $P_f(W)$, is then
\begin{align}
 \hat P_f(u) &\equiv \int P(W)e^{iuW}dW \nonumber\\
 &= \sum_{i,j} e^{iu(\epsilon_j-\epsilon_i)}\Tr[\Pi_j U(\tau)\Pi_i \rho(0)\Pi_i U^\dagger(\tau) \Pi_j] \nonumber\\
 &= \sum_{i,j} \Tr[\Pi_j U(\tau) e^{-iuH(0)}\Pi_i\rho(0)\Pi_i U^\dagger(\tau)e^{iuH(\tau)}\Pi_j] \nonumber\\
 &= \frac{1}{Z(0)}\sum_{i,j}\Tr[U^\dagger(\tau)e^{iuH(\tau)}\Pi_j U(\tau) e^{-iuH(0)}\Pi_ie^{-\beta H(0)}\Pi_i]\nonumber\\
 &= \frac{1}{Z(0)}\sum_i \Tr[e^{iuH_H(\tau)}e^{-(iu+\beta)H(0)}\Pi_i] \nonumber\\
 &= \frac{1}{Z(0)}\Tr[e^{iuH_H(\tau)}e^{-(iu+\beta)H(0)}],
\end{align}
where we have used the completeness relation $\sum_i\Pi_i = \hat I$, and the definition of Heisenberg operator $H_H(\tau) = U^\dagger(\tau) H(\tau) U(\tau)$.
Substituting $u=i\beta$ in the above relation readily gives
\begin{align}
 \av{e^{-\beta W}} = e^{-\beta\Delta F}.
\end{align}

%
%

\section{Implementation of weak measurements}
\label{sec:weak_operators}

The strong measurements are described by the projection operators $\Pi_0$ and $\Pi_1$. However, such measurements completely disrupt the dynamics of the system and collapses it to one of its eigenstates. Furthermore, any real measurement in a laboratory is in general not a projection measurement, but is of a weaker nature where there is no complete collapse of the state after the measurement. Several models of weak measurement have been studied. We will consider the following operator as the one describing a weak measurement \cite{brun}:

\begin{align}
 A(\lambda) &= \left(\sqrt{\frac{1-\tanh\lambda}{2}}\right)\Pi_0 + \left(\sqrt{\frac{1+\tanh\lambda}{2}}\right)\Pi_1.
 \label{weak_operator}
\end{align}
Note that $\lambda\to\pm\infty$ yields the projective measurements $\Pi_1$ and $\Pi_0$. 
The measurement implemented by the above operator was shown to correspond to a random walk along a curve in state space, with the measurement ending when one of the end points is reached, thereby highlighting that any measurement can be generated by weak measurements \cite{brun}. The difference between the weak and the projective measurements is highlighted by the post-measurement state. After the first projective measurement, the state collapses to either $\Pi_0$ or $\Pi_1$. In contrast, under weak measurement, the state $\rho (0)$ (see \ref{initial state}) changes to
\begin{align}
\rho_{\pm}^{'} & = \frac{A(\pm\lambda)\rho(0)A(\pm\lambda)}{\mathrm{Tr}[A(\pm\lambda)\rho(0)A(\pm\lambda)]}\nonumber\\
&=\frac{1}{(a_0^2(\pm \lambda)e^{-\beta \epsilon_0} + a_1^2(\pm \lambda)e^{-\beta \epsilon_1})}\left(a_0^2(\pm \lambda)e^{-\beta \epsilon_0}\Pi_0 + a_1^2(\pm \lambda)e^{-\beta \epsilon_1}\Pi_1 \right), \label{postmeasureweak}
\end{align}
which is a superposition of $\Pi_0$ and $\Pi_1$. Here $a_0 (\lambda) = \sqrt{\frac{1-\tanh\lambda}{2}}$, $a_1 (\lambda) = \sqrt{\frac{1+\tanh\lambda}{2}}$. Hence, as a result of weak measurement, it can be seen that the state does not collapse to the energy eigenstates. As shown in \cite{venkatesh},  measurement operators for which the Crooks  relation must be satisfied for any protocol that connects the initial and the final Hamiltonian within a finite time must be such that that the post-measurement states $|\psi_m (\zeta) \rangle$ have to coincide with the eigenstates $|m;\zeta \rangle$, for $\zeta= 0$ and $\tau$, possibly up to irrelevant phase factors. The post-measurement state Eq. (\ref{postmeasureweak}) clearly does not coincide with eigenstates and hence the fluctuations relations will not be satisfied, in general, under these measurements. This will be seen below explicitly. 

In the limit $\lambda \to \mp \infty$, the post measurement state $\rho_{+}^{'}$, Eq. (\ref{postmeasureweak}), reduces to
\begin{equation}
	\lim_{\lambda \to -\infty}	\rho_{+}^{'}  = \Pi_0,~~\lim_{\lambda \to \infty}	\rho_{+}^{'}  = \Pi_1,
\end{equation}
thereby reducing to the projective states. This is expected as in these limits, the weak measurement operators reduce to projection operators.

\section{Comparison with Known Results}
\label{sec:comparison}

As shown in \cite{venkatesh}, the Crooks Theorem is accurate \emph{only when the measurement operators are projective in nature}. We verify this in the present section, by comparing the left hand and right sides of Eq. \eqref{venkatesh} below, for various strengths of measurement. We choose a Hamiltonian varying linearly with time, and has the form $H(t)=H(0)+Mt$, where
\begin{align}
 H(0) = \begin{pmatrix}
          \epsilon_0 & 0\\
          0 & \epsilon_1
         \end{pmatrix};
         \hspace{1cm}
M = \begin{pmatrix}
          \alpha_0 & 0\\
          0 & \alpha_1
         \end{pmatrix}.
         \label{H0_M}
\end{align}
Here, $\alpha_0$ and $\alpha_1$ are arbitrary constants that perturb the system from its initial equilibrium state, where the intial density matrix was given by
\begin{align}
 \rho(0)=\frac{1}{Z(0)}\begin{pmatrix}
          e^{-\beta \epsilon_0} & 0\\
          0 & e^{-\beta \epsilon_1}
         \end{pmatrix}.
\end{align}

In \cite{venkatesh}, the form of the Crooks theorem consistent with generalized measurements was shown to be:
\begin{align}
 \Tr [U^\dagger(\lambda)Q(u,\tau)U(\lambda)R(u,0)] &= \Tr[U^\dagger(\lambda)R(-u+i\beta,\tau)U(\lambda)Q(-u+i\beta,0)],
 \label{venkatesh}
\end{align}
where
\begin{align}
 Q(u,t)&= \sum_i e^{iu\epsilon_i(t)}A_i^\dagger A_i; \nonumber\\
 R(u,t) &= \sum_j e^{-iu\epsilon_j(t)} A_j e^{-\beta H(t)} A_j^\dagger.
\end{align}
The operators $A_{i,j}$ are defined as $A_+ \equiv A(\lambda)$ and $A_- \equiv A(-\lambda)$.

\begin{figure}[!h]
 \centering
 \begin{subfigure}{0.45\linewidth}
 \includegraphics[width=\linewidth]{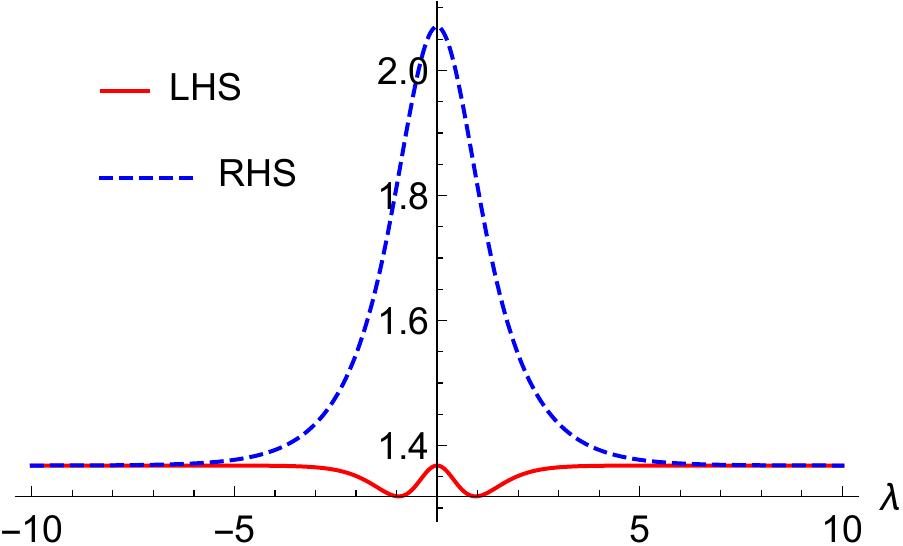}
 \caption{}
 \end{subfigure}
 \hfill
 \begin{subfigure}{0.45\linewidth}
 \includegraphics[width=\linewidth]{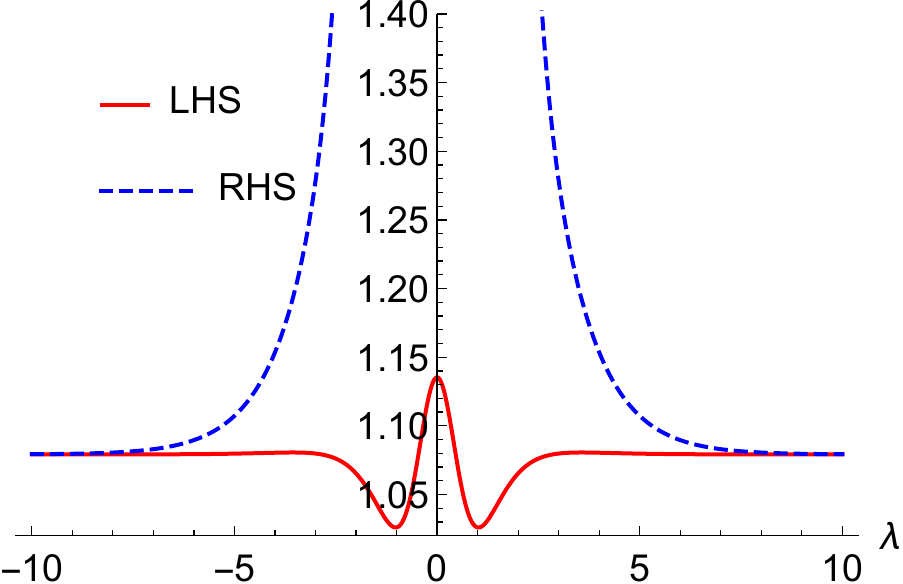}
 \caption{}
 \end{subfigure}
 \caption{(a) Plot of the LHS and RHS of the relation \eqref{venkatesh}, as a function of the measurement strength $\lambda$. The parameters are $\alpha_0=1,~\alpha_1=1,~\beta=1,~u=1,~\epsilon_0=0,~\epsilon_1=1$. (b) Similar plot for a different set of parameters: $\alpha_0=1,~\alpha_1=2,~\beta=2,~u=1,~\epsilon_0=0,~\epsilon_1=1$}
 \label{fig:venk}
\end{figure}

In figure \ref{fig:venk}(a) and (b), we verify this relation for two different sets of parameters (see figure caption). In agreement with the results of \cite{venkatesh}, we find that the left-hand-side and the right-hand-side are exactly equal for projective measurements ($\lambda\to\pm\infty$), whereas for weak measurements, they are not equal. In the weak measurement scenario, the projective operators $\Pi_0=|0\rangle\langle 0|$ and $\Pi_1=|1\rangle\langle 1|$ are replaced by Eq. \eqref{weak_operator}, such that the limits $\lambda\to\pm\infty$ yield the projective measurements $\Pi_1$ and $\Pi_0$. The maximum discrepancy appears for $\lambda=0$, where the two eigenstates of projective measurement contribute equally in the superposed state.

\section{Fluctuation Theorems with weak energy eigenvalues}
\label{sec:new_results}

In \cite{venkatesh}, the derivation of the condition of Crooks Theorem involved the definition of work in terms of the energy eigenstates of the projective (strong) measurements, and likewise the partition function was also defined in terms of these eigenstates. However, since exact energy eigenstates would not be observed under weak measurements, we seek a different definition of work and partition functions in terms of the \emph{weak energy eigenvalues} that will actually be  observed in such measurements.

Next, we seek to find whether a different definition of the partition function leads to a better agreement with the Crooks theorem when the measurement strength is weak.
Since the weak energy values will be returned by the weak measurements, we make an ansatz for the \emph{weak partition function} as (see sec. \ref{sec:temperature} for its practical significance)
\begin{align}
 Z^w(t) &= e^{-\beta \epsilon_+^w(t)} + e^{-\beta \epsilon_-^w(t)}.
 \label{Z}
\end{align}
 Here $\epsilon_{\pm}^w(t)$ are the two \emph{weak eigenvalues} of the Hamiltonian $H(t)$, given by
\begin{align}
\epsilon_\pm^w(t) &= {\rm Tr} \Big[A(\pm \lambda) H \Big]\nonumber\\ &= \left(\sqrt{\frac{1\mp\tanh(\lambda)}{2}}\right)(\epsilon_0+\alpha_0 t) + \left(\sqrt{\frac{1\pm\tanh(\lambda)}{2}}\right)(\epsilon_1+\alpha_1 t).
\end{align}
This is somewhat analogous to the case where quantum discord, a prominent facet of quantum correlations, was studied using weak measurements \cite{superqudiscord}.
Now, in order to evaluate the work distribution (see Eq. \eqref{PfW}), we need to evaluate the path probabilities $P^w(j|i)p_i(0)\equiv \Tr[A_j U(\tau) A_i \rho(0) A_i U^\dagger(\tau) A_j]$, where $i,j=+,-$. These can be easily computed and the results are as follows:
\begin{align}
 P^w(+|-)p_-(0) &= P^w(-|+)p_+(0) = \frac{1}{4}\sech^2(\lambda); \nonumber\\
 P^w(\pm|\pm)p_\pm(0) &=  \frac{1}{4}\sech^2(\lambda)\left[\cosh(2\lambda)\mp\frac{(e^{\beta\epsilon_0}-e^{\beta\epsilon_1})\sinh(2\lambda)}{e^{\beta\epsilon_0}+e^{\beta\epsilon_1}}\right].
\end{align}

The probability density for the weak value of work, $W^w \equiv \epsilon_j^w - \epsilon_i^w$, is given by 
\begin{align}
 P^w_f(W^w) &= \sum_{i,j=+,-} \delta[W^w-(\epsilon_j^w - \epsilon_i^w)] P^w(j|i)p_i^w(0).
 \label{PfW_weak}
\end{align}

\begin{figure}[!ht]
\begin{subfigure}{0.45\linewidth}
 \includegraphics[width=\linewidth]{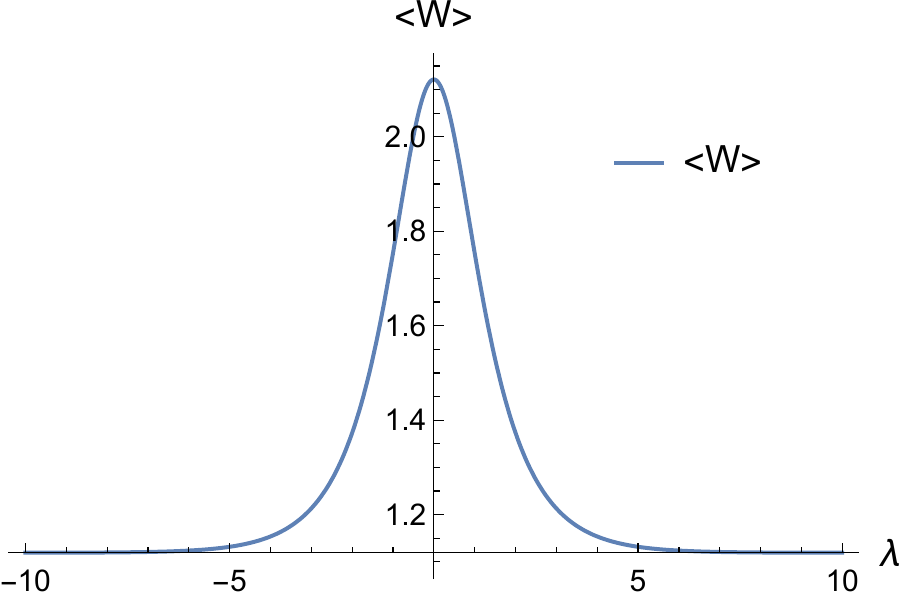}
 \caption{}
 \end{subfigure}
 \hfill
 \begin{subfigure}{0.45\linewidth}
 \includegraphics[width=\linewidth]{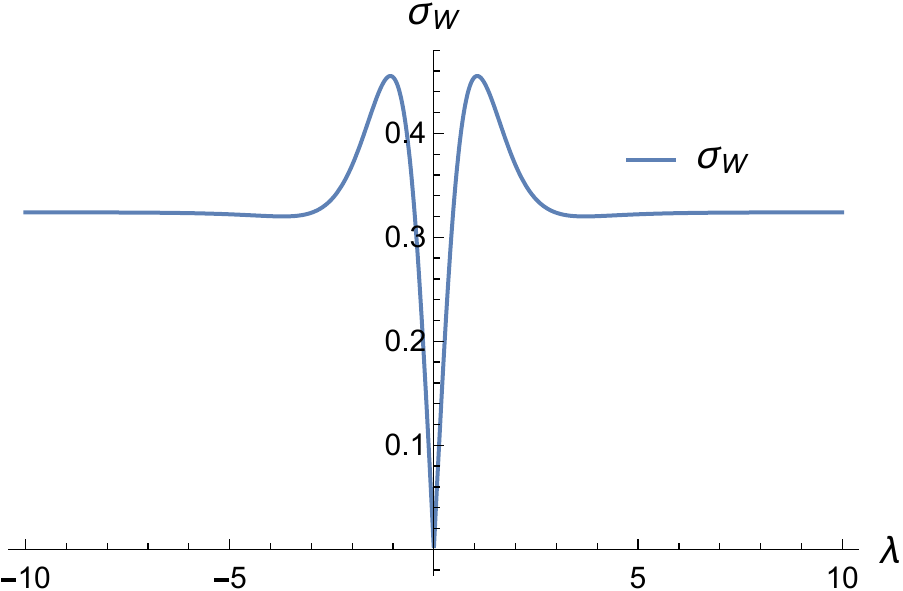}
 \caption{}
 \end{subfigure}
 \caption{(a) Plot of the average work as a function of $\lambda$. (b) Plot of the standard deviation of work as a function of $\lambda$. Parameters are: $\epsilon_0=0,~\epsilon_1=1,~\tau=1,~\alpha_0=1,~\alpha_1=2,~\beta=2$.}
 \label{fig:mean and std}
\end{figure}

We can now find an expression for the mean and variance of work by using Eq. \eqref{PfW_weak}, and the definitions
\begin{align}
 \left<W^w\right> &= \int_{-\infty}^\infty W^wP_f^w(W^w)dW,\nonumber\\
 \sigma_{W^w}^2 &= \int_{-\infty}^\infty (W^w)^2P_f^w(W^w)dW - \left(\int_{-\infty}^\infty W^wP_f^w(W^w)dW\right)^2.
\end{align}
The expression for mean work $\av{W^w}$ can be shown to be
\begin{align}
 \av{W^w} &= \frac{\tau}{2\sqrt{2}}\bigg[\frac{(e^{\beta\epsilon_0}-e^{\beta\epsilon_1})(\alpha_0-\alpha_1)\tanh(\lambda)\left\{\sqrt{1-\tanh(\lambda)}-\sqrt{1+\tanh(\lambda)}\right\}}{e^{\beta\epsilon_0}+e^{\beta\epsilon_1}}\nonumber\\
 &\hspace{4cm}+(\alpha_0+\alpha_1)\left\{\sqrt{1-\tanh(\lambda)}+\sqrt{1+\tanh(\lambda)}\right\}\bigg].
\end{align}
The expression for the variance $\sigma_{W^w}^2$ is very lengthy and unilluminating, and has not been produced here. Instead the plots for the variations of the mean and the standard deviation ($\sigma_W\equiv \sqrt{\sigma^2_W}$) with the weak measurement parameter $\lambda$ have been shown in figure \ref{fig:mean and std}. 
We note that as $\lambda\to 0$, we obtain an equal superposition of states $|0\rangle$ and $|1\rangle$, and the measurement operators $A(\pm\lambda)$ become proportional to identity operator. Thus, the density matrix $e^{-\beta H(0)}/Z(0)$ remains unchanged, and simply undergoes unitary evolution during the process without confronting any interception from measurements.
As a result, the work done during the process is given by $(\alpha_0+\alpha_1)\tau/\sqrt{2}$.
This removes the randomness due to the measurements and leads to a delta-function distribution of work, and hence to the vanishing variance in work distribution in this limit. This limit is also observed to yield the maximum value of mean work. For finite values of $\lambda$, the two measurement operators $A(\lambda)$ and $A(-\lambda)$ are different from each other and retains the randomness of the final weak energy eigenvalues, to a degree that depends on the magnitude of $\lambda$. 
In the limit $\lambda\to\pm\infty$, the two measurement operators reduce to the projection operators $\Pi_1$ and $\Pi_0$, and thus we obtain the results of the usual treatment consisting of projective measurements.

\subsection{Verification of Jarzynski Equality}

The LHS and the RHS of the Jarzynski Equality (JE) are respectively given by
\begin{align}
 \av{e^{-\beta W^w}} &= \int_{-\infty}^\infty e^{-\beta W^w}P_f^w(W^w);\nonumber\\
 e^{-\beta\Delta F^w} &= \frac{Z^w(0)}{Z^w(\tau)}.
 \label{LHS_RHS}
\end{align}
Here $Z^w(t)$ are as given in \eqref{Z}.

\begin{figure}[!ht]
\centering
 \includegraphics[width=0.5\linewidth]{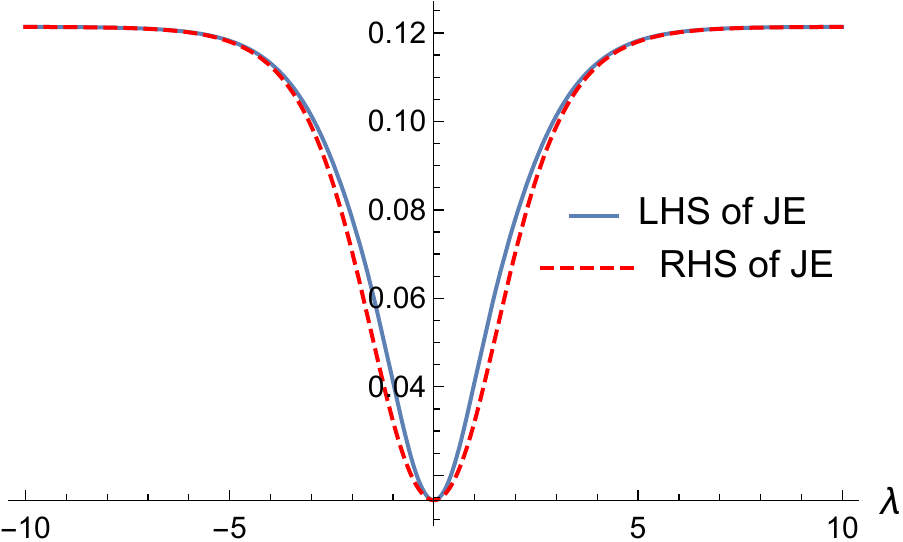}
 \caption{The variation of LHS $\av{e^{-\beta W}}$ and RHS $e^{-\beta\Delta F^w}$ of JE as a function of the measurement strength. Parameters used are: $\epsilon_0=0,~\epsilon_1=1,~\alpha_0=1,~\alpha_1=2,~\beta=2,~\tau=1$.}
 \label{fig:JE}
\end{figure}

The two sides of the Jarzynski equality, given by \eqref{LHS_RHS}, have been  depicted in figure \ref{fig:JE} (the values of the parameters used are given in the figure caption). 
Barring slight deviation in the intermediate parameter regime, we find agreement between the two sides not only for $\lambda\to\pm\infty$, but also for $\lambda$ equal to zero. The agreement in the limit $\lambda\to 0$ can be understood as follows. For the case of the Hamiltonian varying linearly with time, in this limit we have $\epsilon^w_+ = \epsilon_-^w = \frac{1}{\sqrt{2}}(\epsilon_0+\epsilon_1)$, while the weak partition function becomes $Z^w(0)=2\exp\left[-\frac{\beta}{\sqrt{2}}(\epsilon_0+\epsilon_1)\right]$. A similar definition holds at time $t=\tau$ as well, such that one finally obtains
 \begin{align}
  \av{e^{-\beta W^w}} &=  e^{-\beta\Delta F^w} = \exp\left[-\frac{\beta\tau}{\sqrt{2}}(\alpha_0+\alpha_1)\right].
 \end{align}
 %
 This result is in sharp contrast with the approach of \cite{venkatesh}, where the maximum disagreement between the two sides appeared in the $\lambda\to 0$ limit.

\subsection{Verification of Crooks Theorem}

The Crooks Theorem provides a relation between  the work distributions obtained in the forward process, $P_f(W)$, to that along the reverse process, $P_r(W)$. The mathematical statement of the theorem in our case should be 
\begin{align}
 \frac{P_f^w(W^w)}{P_r^w(-W^w)} = e^{\beta(W^w-\Delta F^w)}.
 \label{CFT_weak}
\end{align}
If the forward process involves an external perturbation that varies with time as per the function $\lambda(t)$, then the reverse process can be generated by using the perturbation $\lambda(\tau-t)$. The initial distributions of the forward and reverse processes correspond to the parameter values $\lambda(0)$ and $\lambda(\tau)$, respectively.

One method of verifying the theorem numerically is to directly plot the functions $P_f^w(W^w)e^{-\beta W^w}$ and $P_r^w(-W)e^{-\beta\Delta F^w}$ together and check whether they coincide with each other. But our work distributions are summations of a finite number of delta functions (see eq. \eqref{PfW_weak}), which makes it difficult to plot them directly. Instead, we take the following approach. Suppose we define an arbitrary function $f(W^w)$ of work $W^w$, then multiplying both sides of eq. \eqref{CFT_weak} by $f(W^w)$, we obtain:
\[
 f(W^w)e^{-\beta W^w}P_f^w(W) = f(W^w)e^{-\beta\Delta F^w}P_r^w(-W^w).
\]
An integration over $W^w$ yields
\begin{align}
 \av{f(W^w)e^{-\beta W^w}}_f &= \av{f(-W^w)}_r e^{-\beta\Delta F^w}.
 \label{CFT_verification}
\end{align}
Here, the subscripts $f$ and $r$ on the left and right hand sides of the equation represent averaging over $P_f^w(W^w)$ and $P_r^w(W^w)$, respectively. If this equation remains true for any form of $f(W^w)$, then the Crooks Fluctuation Theorem stands verified.

\begin{figure}[!h]
 \centering
 \begin{subfigure}{0.48\linewidth}
 \includegraphics[width=\linewidth]{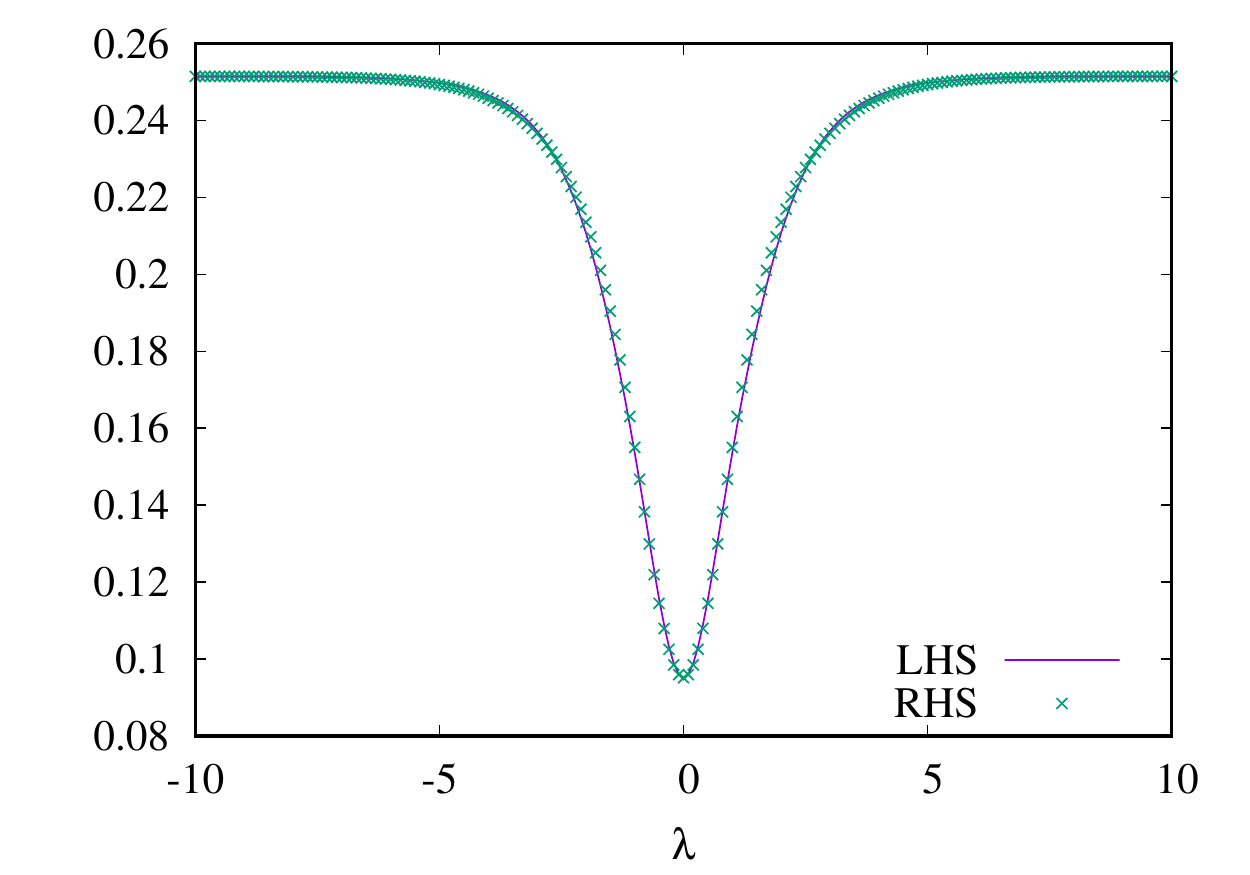}
 \caption{}
 \end{subfigure}
\begin{subfigure}{0.48\linewidth}
 \includegraphics[width=\linewidth]{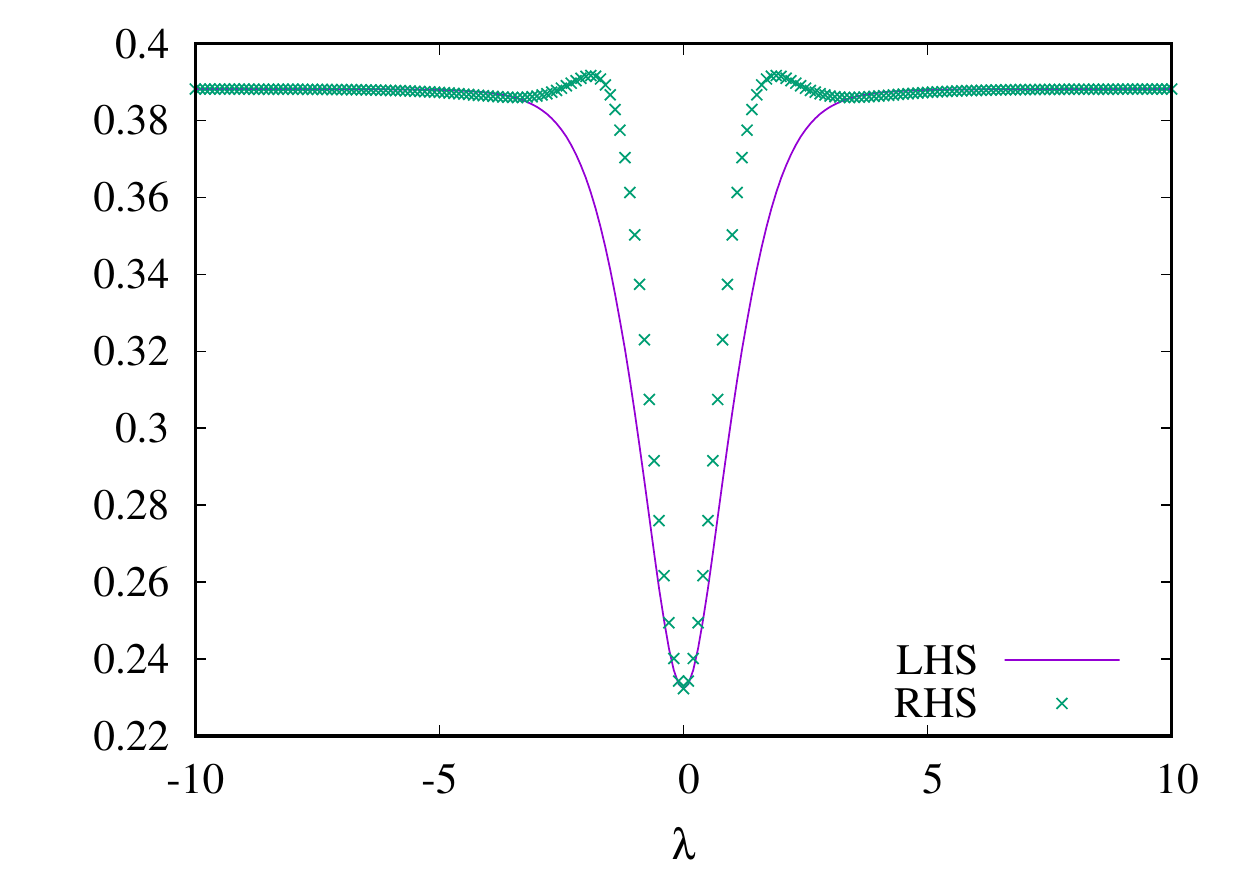}
 \caption{}
 \end{subfigure}
 
 \caption{(a) Verification of Crooks Theorem for $f(W^w)=W^w+(W^w)^2$, with parameter values $\epsilon_0=0,~\epsilon_1=1,~\tau=1,~\alpha_0=1,~\alpha_1=2,~\beta=2$. The purple solid line and the green symbols correspond to the LHS and the RHS of Eq. \eqref{CFT_verification}, respectively. (b) Similar curves for $f(W^w)=W^w+(W^w)^2+(W^w)^3$.}
 \label{fig:CFT_verification}
\end{figure}

Since the forward process is given by $H(t)=H(0)+Mt$, with $M$ given by \eqref{H0_M}, the reverse process can be generated by evolving the system under the Hamiltonian $H(t)=H(0)+M(\tau-t)$. 
 The LHS and RHS of eq. \eqref{CFT_verification} have been shown in figure \ref{fig:CFT_verification}(a), when $f(W^w)$ is chosen to have the following quadratic form: $f(W^w)=W^w+(W^w)^2$. A similar graph for a cubic function of work, $f(W^w)=W^w+(W^w)^2+(W^w)^3$, has been shown in figure \ref{fig:CFT_verification}(b).
We observe that the two sides of the equation deviate for $0<|\lambda|<5$ (deviation is very small for the quadratic $f(W)$ but slightly higher for cubic $f(W)$), and coincide for $\lambda=0$ as well as for larger values of $\lambda$.

\subsection{Measurement of temperature}
\label{sec:temperature}

The notion of inverse temperature $\beta$ in terms of weak value of an operator under weak measurements was recently discussed in \cite{ghosh}, where it was argued that using existing methods of measuring the weak value, an estimate of the (inverse) temperature could be obtained. We show below that using the results of the previous section, it is easy to deduce the temperature of the thermal bath with which the system is in contact at the beginning of either the forward or the reverse process.
We found above that the Crooks theorem for dissipated work approximately holds when the measurements are not projective, when the free energy function are as defined via equations \eqref{Z} and \eqref{LHS_RHS}. Thus we have the following relation:
\begin{align}
 \frac{P_f^w(W^w)}{P_r^w(-W^w)} &\approx e^{\beta (W^w-\Delta F^w)}.
\end{align}
One can rewrite this relation as
\begin{align}
 \beta &\approx \frac{1}{W^w-\Delta F^w}\ln\left[\frac{P_f^w(W^w)}{P_r^w(-W^w)}\right].
\end{align}
Thus, one can find the inverse temperature  by simply using the above relation. The estimate of temperature can be made very precisely when the measurement strengths are either weak or strong, and to a good accuracy for intermediate strengths of the measurements. This can be done, as in the examples discussed above, even when the system is not in contact with the thermal bath (that is, evolving unitarily) throughout the duration of observation $t=0^+$ to $t=\tau$.

\section{Conclusions}

Fluctuation theorems are a set of relations that are valid for a system that  could be far from equilibrium. Such relations have been proved not only for classical systems but also for quantum systems where work is measured by means of two-point projective measurements. We generalized the result to the case where the measurement can have arbitrary strength, with the strongest ones corresponding to projective measurements. The other limit is that of weak measurements, where the state of the system is minimally disturbed and hence the dynamics does not undergo a drastic change under the action of these measurement operators. A recent result shows that the fluctuation theorems for conventional Jarzynski work hold only when the measurement is projective in nature. We verify this result and show that the deviation is maximum when the measurement is weakest. We further generalize this result by defining a weak value of work in terms of weak energy eigenvalues, and show that the work fluctuation theorems are more closely satisfied in this formulation. The deviation from the exact theorems are non-negligible only in a small window of the measurement strength. However, beyond these limits, the deviation becomes imperceptible. We have shown that the approximate fluctuation theorems that we obtain for weak measurements can be used to determine the inverse temperature.


\begin{thebibliography}{99}
	
\bibitem{harbola} M. Esposito, U. Harbola, and S. Mukamel, Rev. Mod. Phys. 81, 1665 (2009).

\bibitem{jarzynski} C. Jarzynski, Annu. Rev. Condens. Matter Phys. 2, 329–351 (2011).

\bibitem{hanggi} M. Campisi, P. Hanggi, and P. Talkner, Rev. Mod. Phys. 83, 771, 2011.

\bibitem{seifert} U. Seifert, {\it Stochastic thermodynamics, fluctuation theorems,
	and molecular machines}, Rep. Prog. Phys. 75, 126001 (2012).

\bibitem{anders} S. Vinjanampathy and J. Anders, Contemporary Physics 57, 545-579 (2016).

\bibitem{rmplinearresponse} U. M. B. Marconi, A. Puglisi, L. Rondoni and A. Vulpiani,   {\it Fluctuation-Dissipation: Response Theory in Statistical Physics}, Rev. Mod. Phys. (2008).

\bibitem{gaspard2007} D. Andrieux and P. Gaspard, J. Stat. Mech. 2007, P02006 (2007).

\bibitem{jarzynski2010} C. Jarzynski, Annu. Rev. Condens. Matter Phys. {\bf 2}, 329 (2010).

\bibitem{jarzynski2007} S. Rahav and C. Jarzynski, J. Stat. Mech. {\bf 2007}, P09012 (2007).

\bibitem{jarzynski2006} V. Y. Chernyak, M. Chertkov and C. Jarzynski, J. Stat. Mech. 2006, P08001 (2006).

\bibitem{seifert2005} U. Seifert, Phys. Rev. Lett. {\bf 95}, 040602 (2005).

\bibitem{seifert2005a} T. Speck and U. Seifert, J. Phys. A {\bf 38}, L581 (2005).

\bibitem{hanggi2009} M. Campisi, P. Talkner and P. H\"anggi, Phys. Rev. Lett. {\bf 102}, 210401 (2009).

\bibitem{hanggi2010} M. Campisi, P. Talkner and P. H\"anggi, Phys. Rev. Lett. {\bf 105}, 140601 (2010).



\bibitem{crooks1} G. E. Crooks, J. Stat. Phys. 90, 1481 (1998).

\bibitem{crooks2} G. E. Crooks, Phys. Rev. E 60, 2721 (1999). 

\bibitem{jarzynski1} C. Jarzynski, Phys. Rev. Lett. 78, 2690 (1997).

\bibitem{monnai} T. Monnai, Phys. Rev. E 72, 027102 (2005).

\bibitem{talknerwork} P. Talkner, E. Lutz, and P. Hanggi, Phys. Rev. E 75, 050102(R) (2007).

\bibitem{paz} A. J. Roncaglia, F. Cerisola, and J. P. Paz, Phys. Rev. Lett. 113, 250601 (2014).

\bibitem{weiss}  U. Weiss, {\it Quantum Dissipative Systems}, Vol. 13 (World Scientific, 2012).

\bibitem{sb}  S. Banerjee, {\it Open Quantum Systems: Dynamics of Non-classical Evolution}, Springer, 2018.

\bibitem{mukamel} M. Esposito, and S. Mukamel, Phys. Rev. E 73, 046129 (2006).

\bibitem{mallick} R. Chetrite and K. Mallick, J. Stat. Phys. 148, 480 (2012).

\bibitem{quan} H. T. Quan and H. Dong, arXiv:0812.4955.

\bibitem{vedral} R. Dorner, S. R. Clark, L. Heaney, R. Fazio, J. Goold and V. Vedral, Phys. Rev. Lett. 110, 230601 (2013).

\bibitem{sekimoto1998} K. Sekimoto, Prog. Theor. Phys. Supp. {\bf 130}, 17 (1998).

\bibitem{siefertstochastic} U. Seifert, Eur. Phys. J. B 64,423 (2008).

\bibitem{sekimoto} K. Sekimoto, {\it Stochastic Energetics}, Springer-Verlag, Berlin, Heidelberg, 2010.

\bibitem{kosloff} R. Kosloff, Entropy 15, 2100-2128 (2013).

\bibitem{lutz} S. Deffner and E. Lutz, Phys. Rev. Lett. 105, 170402 (2010).

\bibitem{jayan} S. Lahiri and A. M. Jayannavar, arXiv:1906.00159.

\bibitem{george}  G. Thomas,  N.  Siddharth,  S.  Banerjee,  and S. Ghosh, Phys. Rev. E 97, 062108 (2018).

\bibitem{yi-kim} J. Yi, and Y. W. Kim, Phys. Rev. E 88, 032105 (2013).

\bibitem{brun2002} T. A. Brun, Am. J. Phys. 70, 719 (2002).

\bibitem{aharanov} Y. Aharonov, D. Z. Albert, and L. Vaidman, Phys. Rev. Lett. 60,
1351 (1988).

\bibitem{brun} O. Oreshkov and T.A. Brun, Phys. Rev. Lett. 95, 110409 (2005).

\bibitem{venkatesh} G. Watanabe, B. P. Venkatesh, P. Talkner, M. Campisi,  and P. Hanggi, Phys. Rev. E 89, 032114, 2014.

\bibitem{superqudiscord} U. Singh, A. K. Pati, Annals of Physics, 343, 141 (2014).


\bibitem{ghosh} A. K. Pati, C. Mukhopadhyay, S. Chakraborty, and S. Ghosh, Phys. Rev. A 102, 012204 (2020).




\end{thebibliography}
\end{document}